\documentclass[pre,preprint,showpacs,showkeys,preprintnumbers,amsmath,amssymb]{revtex4}    
\usepackage{graphicx}% Include figure files  
\usepackage{dcolumn}% Align table columns on decimal point  
\usepackage{bm}% bold math  
\begin{document}  
\newcommand{\be}{\begin{equation}}   
\newcommand{\ee}{\end{equation}} 
\newcommand{\mean}[1]{\left\langle #1 \right\rangle} 
\newcommand{\abs}[1]{\left| #1 \right|} 
\newcommand{\set}[1]{\left\{ #1 \right\}} 
\newcommand{\norm}[1]{\left\|#1\right\|} \newcommand{\eps}{\varepsilon} 
\newcommand{\tNN}{\tilde{\mathbf{X}}_n^{NN}} 
\newcommand{\NN}{\mathbf{X}_n^{NN}} 
\newcommand{\ber}{\begin{eqnarray}} 
\newcommand{\eer}{\end{eqnarray}} 
\newcommand{\lb}{\left(} 
\newcommand{\rb}{\right)}  
\preprint{a} 
 \title{Risk evaluation with enhanced covariance matrix}  
\author{Krzysztof Urbanowicz}  
 \email{urbanow@pks.mpg.de}  
 \affiliation{Faculty of Physics and Centre of Excellence for Complex Systems Research, Warsaw University of Technology\\  Koszykowa 75, PL-00-662 Warsaw, Poland} 
\author{Peter Richmond}  
 \email{richmond@tcd.ie} 
 \affiliation{School of Physics, Trinity College\\  Dublin 2, Ireland} 
\author{Janusz A. Ho{\l}yst}  
 \email{jholyst@if.pw.edu.pl}  
 \affiliation{ Faculty of Physics and Centre of Excellence for Complex Systems Research, Warsaw University of Technology\\  Koszykowa 75, PL-00-662 Warsaw, Poland} 
\date{\today}    \begin{abstract} We propose a route for the evaluation of risk based on a transformation of 
the covariance matrix.  The approach uses a  `potential' or `objective' function. This allows us to rescale 
data from different assets (or sources) such that each data set then has similar statistical properties in terms 
of their probability distributions. The method is tested using historical data from both the  New York and 
Warsaw Stock Exchanges. \end{abstract}  
 \maketitle  \section{Introduction} 

\par  Optimization of portfolios has been much studied since the pioneering work  of Markowitz 
\cite{Markowitz,markowitz2}, who proposed using the mean-variance as a route to portfolio optimization  
\cite{LargeScalePort,distrib,voit,Buchound,Mantegna,Peters,urbnikkei,urbphysa,optimiza,CAPM,optimization,Markowitz,markowitz2,MPS,Poptcorr,Popt2}.  
However the basic construction of the portfolio has not changed much as a result. Computation of Sharp ratios 
\cite{sharp, sharp2} and the Markowitz analysis equate risk with the co-variance matrix.  Portfolio allocations 
are then computed by maximizing a suitably constructed utility function \cite{utility,utility2, utility3}. 
Moreover, the approach taken by Markowitz and many other authors \cite{Markowitz, markowitz2} is essentially 
only appropriate for stochastic processes that follow random walks and exhibit Gaussian distributions \cite{voit,Buchound, Mantegna}.  Many economists 
have sought to use other utility functions and invoke additional objectives \cite{3rdmoment,period} in which 
portfolio weights are computed via maximization of these different utility functionals. Others have introduced 
additional features of the probability distribution such as the third moment or skewness of the returns 
\cite{3rdmoment, period}. This builds in aspects of the deviation of the probability distribution from the 
Gaussian as well as the asymmetry. Introducing even a constant value for the skewness may yield more 
reliable portfolio weights than a calculation in which only the variance or second moment of the distribution 
is used and where the risk of extreme values is seriously underestimated. Similar comments could be made 
about the introduction of the kurtosis which is a first order route to addressing the issue of `fat' tails.

\par In recent years a number of physicists have begun to study the effect of 
correlations on financial risk. Techniques based on Random Matrix 
Theory developed and used in nuclear physics have been applied to reveal the 
linear dependencies between stock market data for both the US and UK markets 
\cite{Buchound, Mantegna}. More recently other workers including one of the present authors have 
used Minimum Spanning Trees methods \cite{plerou, coelho, coelho2} for the same purpose. Spanning 
tree methods seem to yield results that are similar to those obtained using 
random matrix theory but with less effort and the use of less data in the 
sense that only a subset of the correlation matrix is actually used to 
construct the tree. The overall aim, in both cases, is to arrive at optimal 
diversified portfolios. One interesting result obtained in \cite{coelho} was the 
identification of new classifications introduced in the FTSE index ahead of 
their formal introduction by the London authorities. 

%\par In recent years a number of physicists has began to study the effect of correlations on risk. Techniques that invoke for example Random
%Matrix Theory and Minimum Spanning Trees \cite{Buchound, Mantegna, plerou, coelho, coelho2} seek to explore the linear dependencies between stock data and arrive at optimal diversified portfolio.
%{\it It is needed 2 sentences more about econophysics.!!!!}

\par  An important outcome of studies basing on Markowitz approach is the Capital Asset Pricing Model (CAPM) 
\cite{CAPM,CAPMselconsistent, nobel, lim} that relates risk to correlations within the market portfolio 
\cite{CAPM,CAPMselconsistent,nobel,lim} of course the risk now is that all investments will collapse  
simultaneously. Furthermore it is assumed that risk that achieves premiums in the long term should not be 
reducible otherwise arbitrage is possible \cite{CAPMselconsistent}. This is essentially the Arbitrage Pricing 
Theory (APT).  \par  However, key issues remain unresolved. One weakness of CAPM and APT theories 
is that they assume efficiency in the proliferation of market information. In a real market not all investors 
have the same or complete information and arbitrage is possible. Merton \cite{Merton} has  discussed this and 
in so doing has extended CAPM theory to deal more  effectively with small firms for which information is not 
always readily  available.   
\par  Here we concern ourselves with a new approach to the exploitation of 
datasets for the computation of portfolio weights within a diversified portfolio. The method exploits the 
full character of the distribution function for each asset in the portfolio and seeks to maximize the impact of 
correlations. In the next section we discuss the background to our approach and introduce the so-called 
\textit{objective} function. Having established this we show how, from data, we can construct values for a 
renormalised objective function. These are then used in section~\ref{sec3} to obtain both covariance matrices and 
weights for portfolios  of stocks. The calculations are illustrated in section~\ref{sec4} by examples from 
both the US and Warsaw stock exchanges. We also show how the approach modifies the underlying distribution 
of eigenvalues enhancing the  correlations for larger values.  
\section{Objective function\label{sec.2}} \par Consider an 
asset, characterized by a price, $S(t)$ and return $x(t)=\ln S(t+~1)/S(t)$. The objective function, $w(x)$ is 
defined in terms of the stationary probability distribution for returns, $P(x)$, viz: 
\begin{equation} \label{eq1} P(x)=\frac{1}{Z}e^{- w(x)/D}, \end{equation}   where $Z$ is a normalization 
factor. Such functions are familiar to physicists and may be 'derived' by minimizing a 
`free energy' functional, $F(w(x))$, subject to constraints on the mean value of the 'objective' function, 
viz: \begin{equation} \label{eq2} F=\int\limits_{\rm R} {dxP(x)[\ln P(x)+\frac{w(x)}{D}-\lambda } ] \end{equation} 
\par 
Such a form for the probability distribution is also the outcome of a  model that assumes $x$ is 
governed by a generalised Markovian stochastic process of the form 
\begin{equation} \label{eq4} \dot{x}(t)=f(x)+g(x)\varepsilon(t) \end{equation} The Gaussian process, 
$\varepsilon$, satisfies: \begin{equation} \label{eq5} \begin{array}{c}  
\left\langle {\varepsilon (t)\varepsilon (t')} \right\rangle =D \delta  (t-t')\mbox{ } \\    
\left\langle {\varepsilon (t)} \right\rangle =0 \\   \end{array} \end{equation} For the moment 
we leave the form of the functions $f$ and $g$ unspecified  except to say that they only 
depend on $x(t)$. The solution to such a stochastic  process has been deduced elsewhere 
\cite{Zinn,Richmond,Solomon}. Adopting the Ito  convention, the distribution function, $P(x,t)$, 
associated with the process is given  by the Fokker Planck equation: \begin{equation} \label{eq6} 
\frac{\partial P(x,t)}{\partial t}=\frac{\partial ^2}{\partial x^2}\lb \frac{D}{2} g^2(x) P(x,t)\rb-\frac{\partial }{\partial x}\left(  {f(x)P(x,t)} \right) \end{equation} The stationary solution is: 
\begin{equation} \label{eq7} P(x)=\frac{e^{\int {dx\frac{2 f}{(Dg^2)}} }}{{\rm Z}\cdot g^2(x)}=\frac{1}{Z}\exp \lb -\frac{2}{D}\int dx \frac{Dgg'-f}{g^2}\rb \end{equation} Z is a normalization factor.  
\par  A number of different cases are evident as expressed in table~\ref{tab1}.   \begin{table} \caption{\label{tab1} Examples of objective values $w(x)$ and corresponding probability distributions, $P$ for different choices of $f$ and $g$.} \begin{tabular}{|c|c|c|c|} \hline   % after \\: \hline or \cline{col1-col2} \cline{col3-col4} ...   
$f(x)$ & $g(x)$ & $w(x)/D$ & $P(x)\cdot Z$ \\   \hline   $-sgn(x)$ & 1 & $\abs{x}/D$ & $e^{-\abs{x}/D}$ \\   $-x$ & 1& $x^2/D$ & $e^{-x^2/D}$\\   
$\lambda gg'$ & $g(x)\neq const$& $2\lb 1-\lambda/D\rb \ln g$& $\frac{1}{g^{2(1-\lambda/D)}}$\\   $\frac{2x}{\nu}\lb 1+x^2/\nu\rb$ & $1+x^2/\nu$ & $\lb \nu+1\rb/2\ln\lb 1+x^2/\nu\rb$ & $\frac{1}{\lb 1+x^2/\nu\rb^{\lb \nu+1\rb/2}}$\\ \hline \end{tabular} \end{table}   
%KU
Row 4 is obtained from row 3 by choosing $g=(1+x^2/\nu)$, $f=gg'$ and $\lb \nu+1\rb/2=2(1-\lambda/D)$ when we see that the 
distribution function reduces to a Student distribution. 
For $D>0$ we see that $\nu<3$.
On the other hand, we know that Student distribution is defined for $\nu>2$ in order the variance to be finite.
Nevertheless this limitation, that stochastic process can not be defined for $\nu>3$, we can normalize the distribution function,
 because $w(x)$ is well defined for whole spectrum of $\nu>2$ using Eq.~(\ref{eq1}).
%end
In developing our methodology 
in the next sections we shall focus on the use of the student distribution that seems to offer a good fit to the 
data we consider. Tsallis and Anteneodo \cite{Anteneodo} 
have shown how similar multiplicative stochastic processes based on other non-analytic choices for the 
function $f$ and $g$ can lead to q-exponentials. More recently Queiros, Anteneodo and Tsallis \cite{queiros}
have shown that for many financial processes where fat tailed probability functions are empirically observed these Student or Tsallis
distributions are good choices.  
\section{\label{sec3}Portfolio optimization} 
\par As we 
have noted above it is usual for a portfolio of M stocks to compute portfolio weights, $p_{i}$ using the 
covariance matrix, \textbf{C} and defining  the risk, $R$, as: \begin{equation} 
\label{eq10} R=\sum\limits_{i,j} {{\rm {\bf C}}_{i,j} p_i p_j }  \end{equation} Optimizing 
this in the absence of risk free assets yields the weight of stock $i$: \begin{equation} \label{eq11} p_i =\frac{1}{Z}\sum\limits_j {\lb{\rm {\bf C}}^{-1}\rb_{i,j} }  
\end{equation} where $Z=\sum\limits_{i,j} {\lb{\rm {\bf C}}^{-1}\rb_{i,j} }$. 
\par %From our previous discussion, it should be clear that 
%the magnitude of the correlations calculated in this way depend on the relative `objective' values for pairs of stocks, 
%$i$ and $j$. Transforming the objective value will change the magnitude of this correlation. We now assert following 
%the discussion in the previous section that the maximum correlation is obtained by 
It is known that a nonlinear transformation of data can change correlations e.g. correlations of $\abs{x_i}$ decrease much slower
 than $x_i$ \cite{Mantegna}.
We exploit this by introducing a 
particular transformation that increases correlations by
renormalising the objective values 
such that the total set of values, ${x_i(t_j)}$ for all $i$ from 1 to M and $j$ from 1 to N are drawn from a 
common distribution. To effect this change, we first compute 
for each asset the probability distribution by fitting the data for each asset using a student distribution 
characterised by the power law index. We then compute for each value of 
the return $x_i(t_j)$ the corresponding objective value, ${w}_i (x_{t_j})$. These objective values are then transformed 
to yield a set of renormalised objective values as follows: \begin{equation} \label{eq12} 
\tilde{w}_i (x_{t_j } )=w_i (x_{t_j})\frac{\hat{w}}{\bar{w}_i}=w_i (x_{t_j})\frac{\frac{1}{MN}\sum\limits_{i,j}^{M,N} {w_i (x_{t_j } )}}{\frac{1}{N}\sum\limits_j^N  {w_i (x_{t_j } )} }  
\end{equation}  In effect we are renormalising the objective value with its mean value $\bar{w}_i$ 
relative to the overall mean value, $\hat{w}$, of the {\it entire} data set. Having computed these renormalised 
objective values we can now obtain the corresponding set 
of values for $\tilde{x}_{i}(t)$ by inverting the values according to a new student distribution that characterises 
the {\it entire} data set consisting of one value of $\nu$ and MxN values.  
Hence using the result in row 4 of table 1: 
\begin{equation} \label{eq13} \tilde{x}_i(t_j)=\pm \sqrt {\nu (1-e^{2\tilde{w}_i(x_{t_j})/(\nu +1)}} ) \end{equation}  
where $\nu$ is now the tail exponent that characterises the Pdf of the {\it entire} data set.
  
Thus we can compute for our portfolio of M stocks a new covariance  matrix, ${\rm {\bf \tilde {C}}}$ using 
these renormalised values of $\tilde{x}_i(t_j)$. This  yields a new minimized value for the risk: 
\begin{equation} \label{eq14} \tilde {R}=\sum\limits_{k,i=1}^M {{\rm {\bf \tilde {C}}}_{k,i} \tilde {p}_k  \tilde {p}_i }  
\end{equation}    

\section{Illustrative Results and Conclusions\label{sec4}} 

\par
\begin{figure} \begin{center} 
\includegraphics[scale=1]{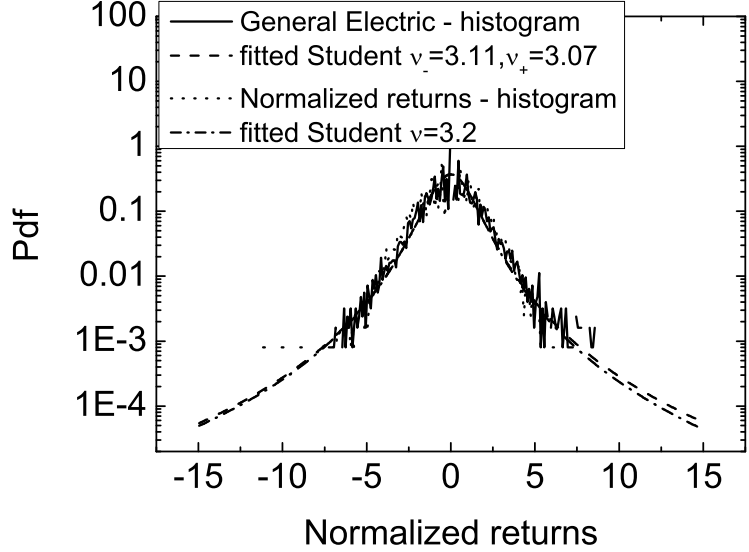} \end{center} \caption{\label{fig.GE} Plot of the histogram of returns and normalized 
returns in the case of General Electric counted in NYSE in years 1999 and 2000 and corresponding Student distributions 
with $\nu_+=3.11$, $\nu_-=3.07$ and $\nu=3.2$ respectively.} 
\end{figure}

\begin{figure} \begin{center} 
\includegraphics[scale=1]{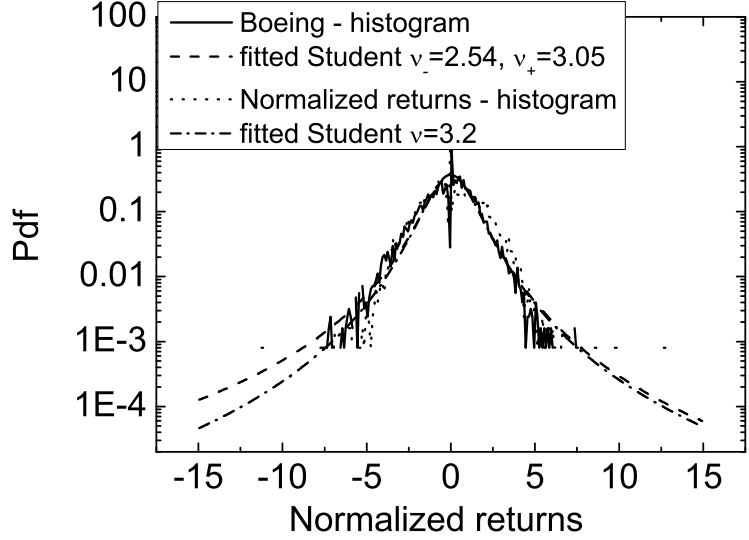} \end{center} \caption{\label{fig.boeing} Plot of the histogram of returns 
and normalized returns in the case of Boeing counted in NYSE in years 1999 and 2000 and corresponding 
Student distributions with $\nu_+=2.54$, $\nu_-=3.05$ and $\nu=3.2$ respectively.} 
\end{figure}

 We show in Figures~\ref{fig.GE} 
and~\ref{fig.boeing} the outcome of implementing the method for a simple portfolio of 2 stocks (i.e, M = 2).  
Specifically we used data for NYSE stocks General Electric and Boeing.  For each stock we used 12500 data points 
extending over the time period January 1999 to December 2000. Student distributions are fitted separately to the 
positive and negative returns. It can be seen that the student distributions for each stock are different prior 
to renormalisation but are the same after renormalisation. The overall changes as a result of our renormalisation 
process are small but we show in figure~\ref{fig3} that they can lead to significant
changes in the distribution of eigenvalues for large eigenvalues. 

\par
\begin{figure} 
\includegraphics[scale=1]{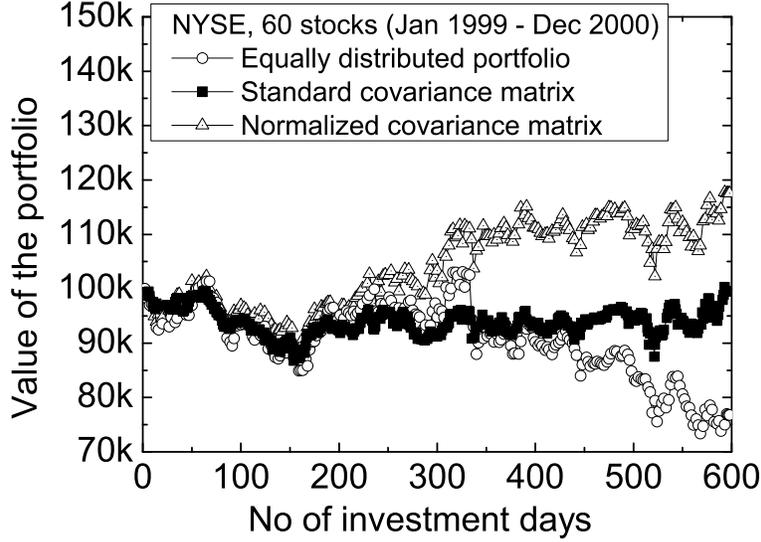}% Here is how to import EPS art 
\caption{\label{fig1} Portfolios runaway of 60 stocks at New York Stock Exchange from  May 1999 to December 2000. 
Equally distributed portfolio (open circles) and portfolio with weights calculated from standard covariance 
matrix Eq.~(\ref{eq11}) (solid squares)  and portfolio with weights calculated from normalized covariance matrix 
are presented.} 
\end{figure}

\begin{figure} 
\includegraphics[scale=1]{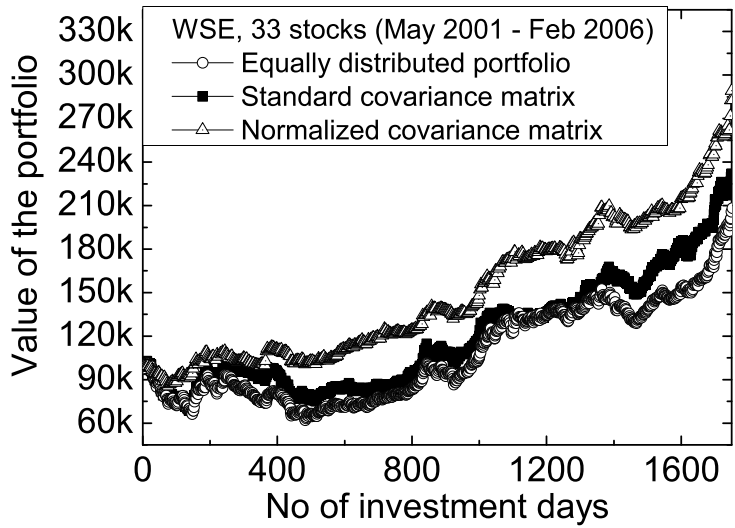}% Here is how to import EPS art 
\caption{\label{fig2} Portfolios 
runaway of 33 stocks at Warsaw Stock Exchange from May 2001 to February 2006. Equally distributed portfolio (open circles) 
and portfolio with weights calculated from standard covariance matrix Eq.~(\ref{eq11}) (solid squares)  and portfolio with 
weights calculated from normalized covariance matrix are presented.} 
\end{figure}

\begin{figure} 
\includegraphics[scale=1]{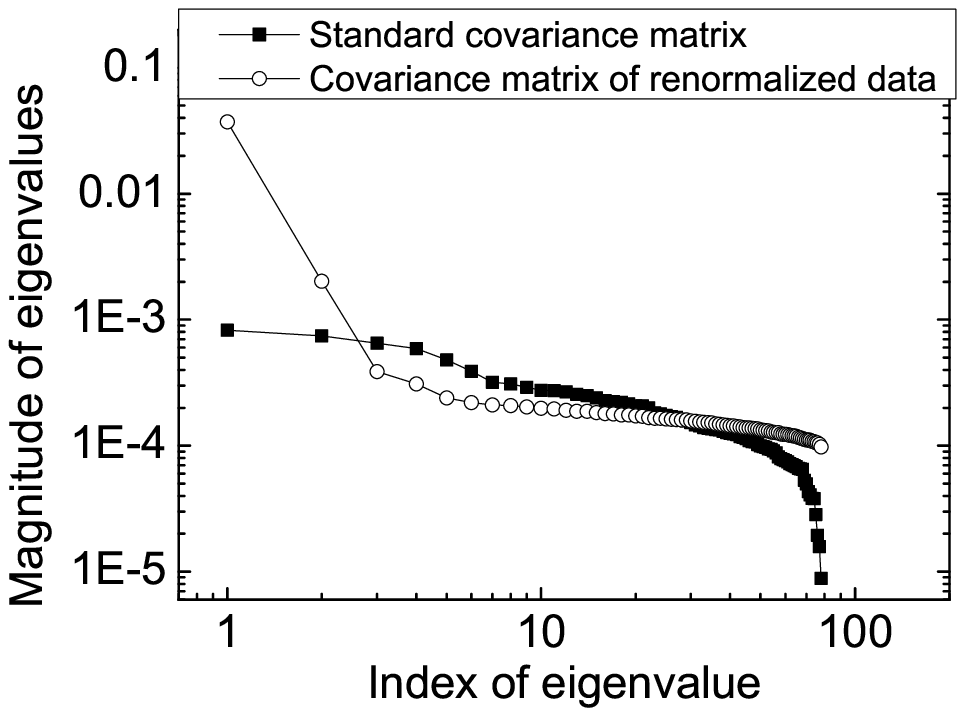}% Here is how to import EPS art 
\caption{\label{fig3} Distribution of eigenvalues of covariance matrices of 78 stocks in NYSE (January 1999 - December 2000). 
Eigenvalues of standard covariance matrix (solid squares) and of covariance matrix from renormalised data (open circles) 
are presented in the graph.} \end{figure}

  We followed up this computation by renormalising data for two different groups of stocks. 
First we selected 60 stocks from the NYSE as before over the period January 1999 to December 2000 and implemented 
the prescription over a moving 75 day window using 1500 points for each window, 
%KU
what corresponds to quarter of hour returns. 
%end
In this way we could compute the 
various elements of the correlation matrix and the associated optimum weights for the different stocks in the 
portfolio as a function of time.  The results are shown in Figure~\ref{fig1}. Figure~\ref{fig2} gives the results 
of a  similar set of calculation for a portfolio of 33 stocks from the Warsaw  stock exchanges over the period 
May 2001 to February 2006. 
% KU added
In order to prevent situations arising where all the money is invested in just one stock we have, in our calculations, imposed the limit $p_i<0.15$. We have checked
that a precise value of thus limit is not crucial for optimization procedure.
% end
\par  Although we have not included transaction costs 
% KU added
(we have changed our portfolio every day, usually by a very small amount),
%end 
in both cases it does seem that 
using data based on our renormalisation procedure we have a route to greater overall returns.  \par  Additional 
insight into the procedure is provided when we compare the distribution of eigenvalues for the standard covariance 
matrix with the corresponding distribution for  the renormalised covariance matrix. These are shown in Figure~\ref{fig3}.  
It can be seen that the transformation procedure enhances correlations as anticipated and this enhancement occurs at 
larger eigenvalues. One could ask why the procedure we have used reduces the risk associated with the portfolio? 
This is because having evaluated the risk connected to each stock then we have a better estimation the weights in 
the portfolio.
%KU
We claim that correlations calculated with standard method underestimate the linear dependencies between stocks, so
the error of the corresponding portfolio risk is much higher. Further, we claim, that we reduce the error related to risk evaluation,
 so risk as a whole is smaller.
%end
 
\par  

It might also be argued at this point that we could by-pass the entire background given in section~\ref{sec.2} and simply fit the 
`best' distribution function to the data as was done, for example, by Levy and Duchin \cite{distrib}. 
Using this approach they obtained
different distributions for different stocks then also obtained different distributions for the same stock at different times.
To our mind this is not a very satisfactory approach and
ignores the evidence from groups led by physicists such as Stanley \cite{Mantegna, plerou} that financial data exhibits 
universal behaviour 
such as scaling, power law tails, etc.
\par Of course an empiricist could still insist that our approach does not yield the best fit and other choices for example
for the entropy might improve our results. To answer this question requires a more extensive study that we have presented here.
\par The covariance matrix is now widely used for the analysis of portfolios. Our 
approach to the exploitation of this matrix that yields new and correct 
linear dependencies clearly has wide application and will, we believe, prove 
to be of considerable benefit to industry practitioners.

  \section{acknowledgment} 
\par This work was supported by the Polish Ministry of Science and Higher Education (Grant No. 134/E-365/SPB/COST/KN/DWM105/2005-2007). Krzysztof Urbanowicz thanks European COST concerted 
action P10 for financial  support that facilitated this collaboration between the Warsaw University of Technology and Trinity 
College Dublin.   
  \end{document}